\documentclass{appolb}
\usepackage{graphicx}

\usepackage{amsmath,amssymb,latexsym}

\begin{document}
\eqsec
\title{Superfluid Picture for Rotating Space-Times}
\author{George Chapline\thanks{\tt chapline1@llnl.gov}
\address{Physics and Advanced Technologies Directorate\\
Lawrence Livermore National Laboratory\\ Livermore, CA 94550,
USA}\\[6mm]
Pawel O. Mazur\thanks{\tt mazurmeister@gmail.com,
mazur@physics.sc.edu}
\address{Department of Physics and Astronomy, University of South Carolina\\
Columbia, SC 29208, USA}}
\maketitle

\begin{abstract}
A new prescription, in the framework of condensate
models for space-times, for physical stationary gravitational
fields is presented. We show that the spinning cosmic string
metric describes the gravitational field associated with the
single vortex in a superfluid condensate model for space-time
outside the vortex core. This metric differs significantly from
the usual acoustic metric for the Onsager--Feynman vortex. We also
consider the question of what happens when many vortices are
present, and show that on large scales a G\"odel-like metric
emerges. In both the single and multiple vortex cases the failure
of general relativity exemplified by the presence of closed
time-like curves is attributed to the breakdown of superfluid
rigidity.
\end{abstract}


\section{Introduction}

The various developments of quantum field theory in curved
space-time have left the false impression that general relativity
and quantum mechanics are compatible. Actually though certain
predictions of classical general relativity such as closed
time-like curves and event horizons are in conflict with a quantum
mechanical description of space-time itself. In particular, a
quantum mechanical description of any system requires a universal
time. In practice, universal time is defined by means of
synchronization of atomic clocks, but such synchronization is not
possible in space-times with event horizons or closed time-like
curves. It has been suggested~\cite{GC92} that the way a global
time is established in Nature is via the occurrence of
off-diagonal long-range quantum coherence in the vacuum state.
This leads to a very different picture of compact astrophysical
objects from that predicted by general
relativity~\cite{CHLS00,MM01,M96,M97,M98,CM0708}.

Certainly, it was historically an unfortunate development, that an unnecessary emphasis
was placed on the high energy-momentum (UV) behavior of scattering amplitudes in `quantum
Einstein gravity', which has blinded many researchers
to the subtleties of the global properties of the gravitational vacuum medium. We have in
mind the infrared (IR) behavior of the gravitational and the Standard Model interactions
of massless elementary excitations in the physically relevant case of the finite positive
vacuum energy density $\epsilon_{\mathrm{vac}}={\mu}^{4}({\hbar c})^{-3}$.
The physical gravitational vacuum state in this case is described by the de Sitter
universe. It was recognized long time ago that the physical gravitational vacuum state is
a highly correlated quantum state of a new kind of matter constituents of which were called
gravitational atoms~\cite{M96,M97,M98,M92}. The fundamental role played by quantum
entanglement in the quantum state of a huge number of strongly interacting bosonic
constituents of the gravitational vacuum medium in the explanation of the underlying
microscopic mechanism responsible for the selection of very small values of the
cosmological constant, and the emergence of gravitational fields described by metric fields
$g_{\mu\nu}$ on space-times, was strongly emphasized
by one of the authors~\cite{M96,M97,M98,M92,M89}. In this letter, we wish to point out the
salient differences between the general relativistic description of rotating space-times
and the picture offered by the assumption that the vacuum state is a quantum condensate.

It has been recognized for a long time that general
relativity fails to describe accurately the physical situation in
the regions of extremely high tidal forces (curvature
singularities) of the type of a Big Bang or the interior of a
black hole. Generally, this failure of general relativity was
considered inconsequential because it was supposed to occur on
Planckian length scales. In this case, a rather soothing philosophy
was adopted to the effect that some mysterious and still unknown
quantum theory of gravitation will take care of the difficulty by
`smoothing out' the curvature singularities. It was recognized
only recently that the physics of event horizons is a second
example of the failure of general relativity but this time on the
macroscopic length scales~\cite{CHLS00,MM01,M96,M97,M98}. In the
following, we consider a third kind of the failure of general
relativity on the macroscopic length scales, associated with the
occurrence of closed time-like curves (CTC). CTCs occur frequently
in\break
\newpage
\noindent analytically extended space-times described by general
relativity once there is rotation present in a physical system
under consideration, which is quite common in Nature.

\section{Quantized vortices in a superfluid and rotating space-times}

The most famous example of a solution to the Einstein equations,
where CTCs occur, is the G\"odel rotating Universe~\cite{Godel49}
though the first example of a rotating space-time with CTCs was
found by Lanczos~\cite{Lanczos24}. In these cases, there is no
universal time because the classical space-time manifold contains
closed time-like curves. G\"odel thought that this indicated
that there was something wrong with the intuitive notion of time
itself. However, in the following, we will show that this strange
behavior can also be viewed as an example of the failure of
classical general relativity on macroscopic length~scales.

As shown in~\cite{CHLS00}, the hydrodynamic equations for a
superfluid that one derives directly from the nonlinear
Schr\"odinger equation are not exactly the classical Euler
equations, but there are quantum corrections to these equations
which become important when a certain quantum coherence length
becomes comparable to length scale over which the superfluid
density varies. One circumstance where this happens is near the
core of a quantized vortex in a rotating superfluid. Although the
physical size of the vortex core in a superfluid is usually small
it can also happen, for example near to the isotropic Heisenberg
point in an $XY$ quantum magnet, that the vortex core has
macroscopic dimensions.

In order to generalize the condensate models of Refs.~\cite{CHLS00,MM01,M96,M97,M98,SUZJUN05} 
to the case of rotating space-times, we
consider the nonlinear Schr\"odinger equation in a general
stationary space-time background described by the line element
\begin{equation}
ds^2 = g_{00}dt^2 + 2g_{0i}dtdx^i + g_{ij}dx^idx^j\,,
\label{eq1}
\end{equation}
where $g_{\mu\nu}$ is time independent. The Lagrangian describing the
condensate of nonrelativistic particles with mass $M$ has the form
\begin{eqnarray}
\textsl{L} &=& \frac{i\hbar
{c^2_{\mathrm{s}}}}{2}g^{00}\left[\psi^{\ast}\left(\partial_t -
g^{ij}g_{oj}\partial_{i}\right)\psi - \psi\left(\partial_t -
g^{ij}g_{oj}\partial_{i}\right)\psi^{\ast}\right]\nonumber\\ 
&& +\frac{\hbar^2}{2M}g^{ij}\partial_i\psi^{\ast}\partial_j\psi +
\frac{\hbar}{2M}g^{0i}{\left(\partial_t\psi^{\ast}\partial_{i}\psi
+ \partial_t\psi\partial_{i}\psi^{\ast}\right)}\nonumber\\ 
&& +\left(\frac{1}{2}M{c^4_{\mathrm{s}}}g^{00} - \frac{1}{2}M{c^2_{\mathrm{s}}} +
\mu\right)\psi^{\ast}\psi - U(|\psi|^2)\,,
\label{eq2}
\end{eqnarray}
where $g^{\mu\nu}$ is the contravariant tensor inverse
to the metric $g_{\mu\nu}$ for the background space-time, $\mu$ is
the chemical potential, $U(|\psi|^2)$ is the interaction potential
energy, and $c_{\mathrm{s}}$ is the velocity of sound in the condensate at
the equilibrium state. The velocity of sound $c_{\mathrm{s}}$ is related to
the interaction potential $U$ by the relation $M{c^2_{\mathrm{s}}} =
|\psi|^2U^{''}(|\psi|^2)$ (and $U'(|\psi|^2) = \mu$, of course).
The equation of motion for the condensate order parameter $\psi$
is
\begin{eqnarray}
i\hbar c^2_{\mathrm{s}}g^{00}\left(\partial_t +
\frac{g^{0i}}{g^{00}}\partial_i\right)\psi &=& {\frac{\hbar^2}{2M}}
\frac{1}{\sqrt{-g}}\partial_i\left({\sqrt{-g}}g^{ij}\partial_j\psi\right)\nonumber\\
&& + \left(U^{'} - \mu\right)\psi -
{\frac{\hbar}{M}}g^{0i}\partial_i\partial_t\psi\,,
\label{eq3}
\end{eqnarray}
where $g$ is the determinant of the spatial metric $g_{ij}$.

It will be useful to write the metric in the form
$g_{\mu\nu} = \eta_{\mu\nu} + h_{\mu\nu}$, where $\eta_{\mu\nu} =
diag({c^2_{\mathrm{s}}}, -1, -1, -1)$. To first order in $h_{\mu\nu}$ the
effect of the background space-time is to introduce a perturbation
$-\frac{1}{2}h^{\mu\nu}T_{\mu\nu}$ in the Lagrangian, where
$T_{\mu\nu}$ is the symmetrized stress-energy-momentum tensor for
the condensate~\cite{SUZJUN05}. Writing $\psi = \sqrt{n}e^{iS}$, where $n =
|\psi|^2$ is the number density of particles in the condensate, we
obtain the velocity field $v_i = \frac{\hbar}{M} \partial_iS$ for
the condensate flow. This representation of $\psi$ leads to the
steady state quantum hydrodynamic equations for $n$ and $v_i$
\begin{eqnarray}
&&\hspace{-6mm}\partial_i\left[n\left(v_i\left(1 - \frac{h_{00}}{2c^2_{\mathrm{s}}}\right) - h_{ij}v_j
-h_{0i}\right)\right] = 0\,,\label{eq4}\\
&&\hspace{-6mm}\frac{\hbar^2}{M\sqrt{n}}\nabla^2\sqrt{n} -
{\frac{\hbar^2}{M\sqrt{n}}}\partial_i\left(h_{ij}\partial_j\sqrt{n}\right) +
2\left(1 - \frac{h_{00}}{2c^2_{\mathrm{s}}}\right)\left(\mu - U^{'}\right)\nonumber\\
&&\hspace{-6mm}- M\left(1 - \frac{h_{00}}{2c^2_{\mathrm{s}}}\right){\vec{v}}^2 + 2Mh_{0i}v_i
- h_{ii}nU^{''}\nonumber\\ 
&&\hspace{-6mm} + Mh_{ij}v_iv_j -
\frac{\hbar^2}{4M}\nabla^2h_{ii} - \frac{\hbar^2}{2Mc^2_{\mathrm{s}}\sqrt{n}}
\vec{\nabla}\cdot\left(h_{00}\vec{\nabla}\sqrt{n}\right)
= 0\,.\label{eq5}
\end{eqnarray}
Our philosophy in the following will be to find the
classical metrics which produce superfluid flows with vortices
when $h_{00} = 0$ and $\partial_{_3}g_{\mu\nu} = 0$. The metric in
our action is not a dynamical field. Instead, the metric components
only act as Lagrange multipliers. The role of these Lagrange
multipliers is to enforce the local equilibrium in the condensate.
The homogeneous vacuum state of the condensate is characterized by
$|\psi| = {\mathrm{const.}}$, $g_{\mu\nu} = \eta_{\mu\nu}$ and $U^{'} = \mu$.

We first seek a solution of Eqs.~(\ref{eq4}) and (\ref{eq5})
corresponding to a single vortex in the condensate. The phase $S$
of the condensate corresponding to a single vortex has a simple
form: $S = N\varphi$, where $\varphi$ is the azimuthal angle
defined by the formula $\varphi = {\mathrm{Arctan}}(\frac{x^2}{x^1})$ and $N$
is the vortex number which is an integer. The velocity field
corresponding to the vortex configuration~is
\begin{equation}
v_i = N\frac{\hbar}{M} \partial_i\varphi\,.
\label{eq6}
\end{equation}
It is convenient to use the following well known
relation (here, the indices $i, j$ take values 1 and 2)
\begin{equation}
\partial_i\varphi = -\epsilon_{ij}\partial_j{\mathrm{ln}}r\,,
\label{eq7}
\end{equation}
which yields
\begin{equation}
v_i = - {N\kappa\over {2\pi r^2}}\epsilon_{ij}x_j\,,
\label{eq8}
\end{equation}
where $\kappa = \frac{h}{M}$ is the fundamental unit of
quantized circulation $\oint \vec{v}\cdot
d\vec{l}$ or the flux of the vorticity field
$\omega_{ij} = \partial_iv_j - \partial_jv_i$. 
The velocity field of a vortex $v_i$ has the form of the
Aharonov--Bohm electromagnetic potential~\cite{AB59}, while the vorticity 
$\omega= \frac{1}{2}\epsilon_{ij}\omega_{ij} =
\epsilon_{ij}\partial_iv_j$ is an analog of the Aharonov--Bohm
magnetic field produced by an infinitely thin solenoid, $\omega =
\kappa\delta(x^1)\delta(x^2)$.

It turns out that because of the presence of the
potentials $h_{0i}$ and $h_{ij}$ in the hydrodynamic equations, the
superfluid density $n$ will be nearly constant when $r$ is greater
than the coherence length $\xi = \frac{\hbar}{M c_{\mathrm{s}}}$. Indeed, it
is straightforward to show that if $n$ is constant and the
velocity has the form given in Eq.~(\ref{eq8}), then Eqs.~(\ref{eq4}) and (\ref{eq5}) have
a solution
\begin{eqnarray}
N &=& 1\,,
\label{eq9}\\
h_{00} &=& 0\,, \qquad  h_{0i} = v_i\,,
\label{eq10}\\
h_{ij} &=& - \frac{v_iv_j}{c^2_{\mathrm{s}}}\,.
\label{eq11}
\end{eqnarray}
These values for the potentials $h_{\mu\nu}$ are
equivalent to the metric for the background space-time of the
local `spinning cosmic string' solution of the Einstein field
equations~\cite{Maz1,Maz2,Maz3} (see also~\cite{Star63,DJH84,dSJack89}) 
in the region where $n$ is constant;
\ie for $r \gtrsim \xi$. The line element for this solution (for
$r > 0$) has the form~\cite{Maz1,Maz2,Maz3}
\begin{equation}
ds^2 = (c_{\mathrm{s}} dt + Ad\varphi)^2 - dr^2 - r^2d{\varphi}^2 - dz^2\,,
\label{eq12}
\end{equation}
where $A = {\kappa\over {2\pi c_{\mathrm{s}}}} = \xi$. The
string-like singularity at $r=0$ has neither mass density nor
pressure, so space-time is flat for $r > 0$. However, the string
rotates resulting in frame dragging. This frame dragging is
represented by the appearance of a vector potential $A_i$~\cite{Maz1,Maz2,Maz3} 
with azimuthal component $A_{\varphi} = A ={\kappa\over {2\pi c_{\mathrm{s}}}}$. 
The frame dragging implied by the
metric (\ref{eq12}) is evidently closely related to the velocity field
surrounding a single vortex filament in a superfluid. Indeed, De
Witt pointed out some time ago~\cite{Bryce66} that the vector
potential $A_i$ associated with frame dragging can be formally
identified as the vector potential for a superconductor. Kirzhnits
and Yudin~\cite{Kirz95} have also studied stationary superfluid
flows in the presence of gravitational fields $g_{0i}$ produced by
rotating compact, massive objects (superfluid cores of\break
\newpage
\noindent  neutron
stars). Balasin and Israel~\cite{BalasinIsrael99} have concluded
that vortex filaments in a superfluid neutron star do produce
gravimagnetic forces, contrary to the statements in the
literature.

It should be noted that excitations other than
collective bosonic excitations in the condensate, for example
massless (massive) fermionic and bosonic excitations or
impurities, will feel the gravitational field, Eq. (\ref{eq12}),
associated with the vortex. However, this field is not the same as
the acoustic metric seen by the condensate excitations. The
scattering cross-section for fermionic (bosonic) particles will be
given by the Aharonov--Bohm cross-section~\cite{Maz2,Maz3} as is
the scattering of quasiparticle excitations of unit electric
charge on Abrikosov vortices~\cite{Abrikosov57} in type II
superconductors. In this sense, the `spinning cosmic string' is a
gravitational analog of the Abrikosov vortex~\cite{Maz2,Maz3}.
This is also the reason why one of the authors has called the
scattering of relativistic particles by gravitational vortices the
gravitational Aharonov--Bohm effect~\cite{Maz2,Maz3}. The
scattering cross-section for condensate excitations has been given
in Ref.~\cite{Grisha98} and for the reasons just mentioned is not
the same as the gravitational Aharonov--Bohm scattering
cross-section~\cite{Maz2,Maz3}.

The space-time corresponding to the metric (\ref{eq12}) does
not have a universal time because closed time-like curves appear
close to the axis of the gravitational vortex. What does not seem
to have been noted before, though, is the fact that closed
time-like curves appear in the gravitational vortex background
(\ref{eq12}) at exactly the radius, where a classical hydrodynamic
description of the superfluid begins to fail. Indeed, the
superfluid velocity (\ref{eq8}) will become comparable to the velocity of
sound $c_{\mathrm{s}}$ when the radius $r$ is close to the quantum
coherence length $\xi$. Therefore, superfluid rigidity and
classical hydrodynamics break down as one enters the core of the
vortex. Remarkably, this breakdown of a classical description of
the superfluid seems to be closely related to the breakdown of
causality in classical GR associated with the formation of closed
time-like geodesics. The condition for the appearance of closed
time-like curves in a rotating space-time is that
$g_{\varphi\varphi} > 0$, which for the gravitational vortex
metric (\ref{eq12}) becomes the condition
\begin{equation}
r < r_{\mathrm{c}} = {\kappa\over 2\pi c_{\mathrm{s}}} = \xi\,.
\label{eq13}
\end{equation}
That is, closed time-like curves appear in the
gravitational vortex solution of the Einstein equations near to
the axis of the string where the velocity of frame dragging
exceeds the speed of light. In the superfluid picture, this
corresponds to the core of the vortex where the superfluid flow
velocity exceeds the speed of sound $c_{\mathrm{s}}$. As previously
discussed, this is just where a classical hydrodynamic description
of the fluid flow in a quantized superfluid vortex breaks down.
Indeed, the solution to the equations of quantum hydrodynamics in
the presence of the potentials $h_{\mu\nu}$ given by Eqs.~(\ref{eq10}),
(\ref{eq11}) is valid only in the region where the condensate particle
density $n$ is constant. The corresponding space-time metric~(\ref{eq12})
is perfectly well behaved in this region ($r > \xi$). It is only
after the na\"ive analytic continuation of the metric (\ref{eq12}) to
the region $r < \xi$ is attempted that the causality violating
regions appear in the space-time.

This observation provokes one to ask if the
appearance of closed time-like curves in solutions of the
classical Einstein field equations might always be associated with
a breakdown of superfluid rigidity. In particular, one might
wonder if the appearance of closed time-like curves in
G\"odel-like universes is related to the behavior of rotating
superfluids. The G\"odel metric for a rotating universe can be
written in the form~\cite{Reboucas83}
\begin{equation}
ds^2 = (cdt +  {\mit\Omega}(r)d\varphi)^2 - dr^2 - f^2d\varphi^2 - dz^2\,,
\label{eq14}
\end{equation}
where ${\mit\Omega}(r) =
\frac{4{\mit\Omega}}{m^2}{\mathrm{sinh}}^2(\frac{mr}{2})$ and $f(r) =
\frac{1}{m}{\mathrm{sinh}}(mr)$. In the limit of small~$r$, ${\mit\Omega}(r)$
approaches ${\mit\Omega} r^2$. The off-diagonal metric component
$g_{0\varphi}$ equals the velocity potential inside a body rigidly
rotating with angular velocity ${\mit\Omega}$. It can be seen that the
metric component $g_{0\varphi}$ for the G\"odel universe has a
very different dependence on radius from that of the gravitational
vortex. However, as we shall now see this very different behavior
is characteristic of what happens in a rapidly rotating
superfluid.

Feynman pointed out~\cite{Feynman55} that when many
vortices are present, the velocity of rotation in the superfluid
will approach that of a rigidly rotating body; \ie $\vec{v} =
\vec{\mit\Omega}\times\vec{r}$. When the area density $\sigma$ of
vortices is not too high, it is reasonable to approximate the phase
in Eq.~(\ref{eq6}) as a sum $S = \sum_a {\mathrm{Arg}}(w - w_a)$, $w = x^1 + ix^2$
of phases of individual vortices each with vortex number $N = 1$.
Using Eq.~(\ref{eq7}), the velocity field in this approximation can be
written in the form
\begin{equation}
v_i = \frac{\kappa}{2\pi}\partial_iS = -
\frac{\kappa}{2\pi}\epsilon_{ij}\partial_j\sum_a{\mathrm{ln}}\left|\vec{x}
- \vec{x_a}\right|\,.
\label{eq15}
\end{equation}
Evaluating the vorticity $\omega =
\epsilon_{ij}\partial_iv_j$ and replacing the sum in Eq.~(\ref{eq15}) by
an integral, we obtain
\begin{equation}
\omega = \frac{\kappa\sigma}{2\pi}\nabla^2_x\int d^2y
{\mathrm{ln}}\left|\vec{x} - \vec{y}\,\right|\,.
\label{eq16}
\end{equation}
Using the relation $\nabla^2_xln|\vec{x} -
\vec{y}\,| = {2\pi}\delta^{(2)}(\vec{x} -
\vec{y}\,)$, we obtain $\omega = \kappa\sigma$. It follows
from Eqs.~(\ref{eq15}) and (\ref{eq16}) that $v_i = -
\frac{\kappa\sigma}{2}\epsilon_{ij}x_j$ which means that this
velocity field is indeed that of a rigid body rotating with the
angular velocity ${\mit\Omega} = \frac{\kappa\sigma}{2}$.
\newpage
Since the gravitational vortex solution (\ref{eq12}) is
spatially flat, it makes sense to construct a new solution to the
Einstein equations by simply superposing the velocity fields, Eq.
(\ref{eq15}), corresponding to a collection of parallel gravitational
vortices. Following the same line of reasoning that leads one to
rigid body rotation in the case of many superfluid vortices, one
would surmise, based on the identification $h_{0i} = v_i$, that in
the presence of many gravitational vortices the metric of
space-time would assume the form
\begin{equation}
ds^2 = \left(c_{\mathrm{s}}dt +  \frac{1}{c_{\mathrm{s}}}{\mit\Omega} r^2 d\varphi\right)^2 - dr^2 -
r^2d\varphi^2 - dz^2\,.
\label{eq17}
\end{equation}
This metric is, in fact, just the Som--Raychaudhuri
solution of the Einstein field equations~\cite{Reboucas83,SomRaychaudhuri68}. 
This metric can be obtained
from the G\"odel metric Eq.~(\ref{eq14}) by letting $m\rightarrow 0$. It
can be seen that the velocity of frame dragging for the metric
(\ref{eq17}) is just the velocity inside a rigidly rotating body. The
condition for the appearance of closed time-like curves, \ie
\mbox{$g_{\varphi\varphi} > 0$,} in the Som--Raychaudhuri space-time is
\begin{equation}
{\mit\Omega} r_{\mathrm{c}} > c_{\mathrm{s}}\,.
\label{eq18}
\end{equation}
That is, closed time-like curves appear when the
velocity of frame dragging exceeds the speed of light. In contrast
with the gravitational vortex, closed time-like curves appear in
the Som--Raychaudhuri space-time at large radii. The appearance of
closed time-like curves in G\"odel space-times mimics the
behavior of Som--Raychaudhuri space-time in that the closed
time-like curves appear at large radii. In particular, for the
G\"odel metric (\ref{eq14}) the condition for the appearance of closed
time-like curves is
\begin{equation}
\frac{2{\mit\Omega}}{m}{\mathrm{tanh}}\frac{mr_{\mathrm{c}}}{2} > c_{\mathrm{s}}\,.
\label{eq19}
\end{equation}
When $m\rightarrow 0$, this condition reduces to Eq.~(\ref{eq18}). 
When $m = 2{{\mit\Omega}\over c}$, the radius where the velocity of
frame dragging approaches the speed of light recedes to infinity,
and the space-time will be free of closed time-like curves
everywhere. We now wish to inquire as to the significance of the
conditions (\ref{eq18}) and (\ref{eq19}) from the point of view of a rotating
superfluid. Evidently then, a superfluid description for the
metrics (\ref{eq14}) and (\ref{eq17}) will require an external rotating container
of normal matter to create a frame dragging potential. The
elementary fact that this container cannot rotate faster than the
speed of light leads to the conditions (\ref{eq18}) and (\ref{eq19}). The
occurrence of solid body-like frame dragging in the G\"odel and
Som--Raychaudhuri metrics may seem to be incompatible with a
superfluid interpretation for space-time because
$\vec{\nabla}\times\vec{v} = 2\vec{\mit\Omega}$ for a solid body
rotating with angular velocity $\vec{\mit\Omega}$, whereas the flow
velocity of a superfluid must have zero curl since it is the
gradient of a phase. The resolution of this paradox is that the
solid body rotation curve corresponds to a coarse-grained average
of the velocities from an array of individual vortices. The
superfluid as a whole will respond to the frame dragging created
by the array of vortices leading to the G\"odel-like metrics. In
between the vortices, the flow is irrotational so
$\vec{\nabla}\times\vec{v} = 0$ in the superfluid condensate.

In contrast with the case of a single vortex, the
coarse-grained potentials associated with the array of vortices do
not satisfy the time independent hydrodynamic Eqs.~(\ref{eq4}) and (\ref{eq5}).
Indeed, in contrast with the case of the single vortex, the term
$\nabla^2 h_{ii}$ in Eq. (\ref{eq5}) which comes from the quantum pressure
no longer cancels the term $h_{ij} v_i v_j$ which arises as a
relativistic correction to the kinetic energy density of the
condensate. Although a simple superposition, Eq.~(\ref{eq15}), of the
single vortex solution, Eqs.~(\ref{eq9})--(\ref{eq11}), does not satisfy the
superfluid Eqs. (\ref{eq4})--({\ref{eq5}), there do exist multi-vortex solutions. In
particular, there exist time independent solutions representing a
regular lattice of vortices, the Tkachenko lattice~\cite{Tkachenko}.

When an impulse of energy is applied to a very low
temperature rotating superfluid condensate, then a turbulent state
containing a time dependent tangle of quantum vortices can develop~\cite{Vinen00}. 
Such a regime is known as quantum turbulence. If
space-time is indeed a condensate and the conditions for the
development of quantum turbulence, \ie rotation and an impulse
of energy are met, then there should be characteristic
observational signatures. For example, the onset of quantum
turbulence in cosmological space-times would lead to a
characteristic scale-free spectrum of energy density fluctuations.

\vspace{7mm}
The authors would like to thank their colleagues Yakir Aharonov, James Bjorken, 
Robert Laughlin, Emil Mottola, David Santiago and Andrzej\break Staruszkiewicz for valuable
comments. One of us (G.C.) would like to acknowledge hospitality
extended to him at the University of South Carolina where this
paper has been completed in March 2004. This material is based upon work
(partially) supported by the National Science Foundation under
Grant No. 0140377 (P.O.M.). This work was also performed (in
part) under the auspices of the U.S. Department of Energy by
University of California Lawrence Livermore National Laboratory
under contract No. W-7405-Eng-48 (G.C.).

\newpage

\noindent{\bf Note added in proof:}\\

The last paragraph of this paper was made very short for brevity reasons because the paper
was initially formatted as a letter and thus no formulae were given to support the claim
that quantum turbulence
in the context of the superfluid model of gravitational fields/space-times leads `... to a
characteristic scale-free spectrum of energy density fluctuations.'

In an earlier unpublished work of one of the authors~\cite{POM87}, the correlation
functions of the statistical random velocity field $v_i({\bf x})$ in the three-dimensional
conformal field theory (3D CFT) describing quantum turbulence on large scales
(and the `normal' fluid $K41$ turbulence; the celebrated Kolmogorov--Obukhov $\frac{5}{3}$
law) were computed
\begin{equation}
\left\langle v_i({\bf x})v_j({\bf y})\right\rangle=
P_{ij}({\bf x},{\bf y};{\Delta}_{v})\left|{\bf x}-{\bf y}\right|^{-2{\Delta}_{v}}\,,
\label{eq20}
\end{equation}
where ${\Delta}_{v}$ is the scaling dimension of the statistical (random) velocity
field $v_i({\bf x})$, and from the condition of the vanishing divergence of the velocity
field (outside the vortex cores of quantized vortices) $\partial_i v_i=0$, we compute the
3D symmetric tensor $P_{ij}({\bf x},{\bf y};{\Delta}_{v})$
\begin{equation}
P_{ij}({\bf x},{\bf y};{{\Delta}_{v}}) = 
C[(1-{\Delta_{v}})\delta_{ij}+\Delta_{v}n_in_j]\,,
\label{eq21}
\end{equation}
where $C$ is a constant and
\begin{equation}
n_i=n_i({\bf x},{\bf y})=\frac{({\bf x}-{\bf y})_i}{|{\bf x}-{\bf y}|}\,.
\label{eq22}
\end{equation}
At large distances, the line integral of the statistical velocity field
$\oint_{_{\Gamma}(R)} v_idx_i$ along the closed contour ${\Gamma}(R)$, 
which is a circle of radius $R$, scales like $R^0$ because of the 
Onsager--Feynman quantization condition~\cite{Feynman55}. 
This means that on the very large distance scale when the contour
encircles the tangle of quantized vortices the scaling dimension
${\Delta}_{v}$ of the statistical velocity field $v_i({\bf x})$ is equal to one:
${\Delta}_{v}=1$.

The na\"ive scaling dimension of the kinetic energy $\epsilon$ is ${\Delta}_{\epsilon}=2$.
The composite statistical operator $\epsilon({\bf x})=\frac{1}{2}\rho
v_i({\bf x})v_i({\bf x})$, with $\rho={\mathrm{const.}}$, scales with the lowest scaling dimension
${\Delta}_{\epsilon}=2{{\Delta}_{v}}$. This translates to the correlations of the composite
energy density operator $\epsilon({\bf x})$ that is equivalent to the Zel'dovich--Harrison
scaling spectrum of the energy density fluctuations.

In fact, the experiments on quantum turbulence in superfluid $^4$He reported during the
COSLAB Workshop in Bilbao in July 2003 showed that at large distance scales velocity
correlations display the scaling dimension ${\Delta}_{v}=1$. This behavior translates to the
Zel'dovich--Harrison power spectrum for energy density $P_{\epsilon}(k)\sim k^{n}$, where
$k=|{\bf k}|$ and $n=1$. Recall that the power spectrum $P_{\epsilon}(k)$ is defined by
the Fourier transform of the two-point correlation function
$\langle{\epsilon({\bf x})\epsilon({\bf y})}\rangle$
by the formula
\begin{equation}
\langle{\epsilon({\bf k})\epsilon({\bf k^{'}})}\rangle
=\delta^{(3)}({\bf k}+{\bf
k^{'}})P_{\epsilon}(|{\bf k}|)\,.
\label{eq23}
\end{equation}
In the case of the two-point correlation function with scaling
\begin{equation}
\langle{\epsilon({\bf x})\epsilon({\bf y})}\rangle 
\sim |{\bf x}-{\bf y}|^{-2{\Delta}_{\epsilon}}\,,
\label{eq24}
\end{equation}
the power spectrum $P_{\epsilon}(k){\sim k^n}$, where the exponent
$n=2{{\Delta}_{\epsilon}}-3$ is entirely given in terms of the scaling dimension
${\Delta}_{\epsilon}$. Indeed, it was shown~\cite{AMM97} that the na\"ive scaling of the
energy density fluctuations in the `early universe' (${\Delta}_{\epsilon}=2$) corresponds to
the celebrated Zel'dovich--Harrison power spectrum with the exponent $n_{\mathrm{ZH}}=1$. There is room for `anomalous dimensions' though~\cite{POM87,AMM97}.

\flushleft

\end{document}